\documentclass[12pt]{article}
\usepackage[margin=1in]{geometry}
\usepackage{comment} 
\usepackage{lipsum} 
\usepackage[dvips]{graphicx}

\usepackage{xcolor}
\usepackage{times}
\usepackage{epstopdf}
\usepackage{cite}
\usepackage{amsmath,amsfonts,amssymb,amsthm,epsfig,epstopdf,titling,url,array}
\usepackage{mathtools , nccmath}
\usepackage{float}
\usepackage{verbatim}
\usepackage{nicefrac}

\usepackage [justification=centering , margin=2cm , labelsep=period,font=small,labelfont=bf]{caption}
\usepackage{booktabs}
\usepackage{tabularx}
\usepackage{subcaption}
\usepackage{afterpage}
\usepackage{placeins}
\usepackage{algorithm}
\usepackage[noend]{algpseudocode}
\newcolumntype{H}{>{\setbox0=\hbox\bgroup}c<{\egroup}@{}}

\providecommand{\keywords}[1]{\textbf{\textit{Key words---}} #1}

\makeatletter
\def\BState{\State\hskip-\ALG@thistlm}

\DeclarePairedDelimiter{\nint}\lfloor\rceil

\usepackage{authblk}

\title{A Weight-based Information Filtration Algorithm for Stock-Correlation Networks}
\author{Seyed Soheil Hosseini}
\author{Nick Wormald \thanks{Research supported by  ARC DP160100835.} }
\author{Tianhai Tian}
\affil{School of Mathematics, Monash University} 
\date{\vspace{-5ex}}

 \def\@testdef #1#2#3{%
   \def\reserved@a{#3}\expandafter \ifx \csname #1@#2\endcsname
  \reserved@a  \else
 \typeout{^^Jlabel #2 changed:^^J%
 \meaning\reserved@a^^J%
 \expandafter\meaning\csname #1@#2\endcsname^^J}%
 \@tempswatrue \fi}

\begin{document}

\maketitle

\begin{abstract}
Several algorithms have been proposed to filter information on a complete graph of correlations across stocks to build a stock-correlation network. Among them the planar maximally filtered graph (PMFG) algorithm uses $3n-6$ edges to build a graph whose features include a high frequency  of small cliques and a good clustering of stocks. We propose a new algorithm which we call proportional degree (PD) to filter information on the complete graph of normalised mutual information (NMI) across stocks. Our results show that the PD algorithm produces a network showing better homogeneity with respect to cliques, as compared to economic sectoral classification than its PMFG counterpart. We also show that the partition of the PD network obtained through normalised spectral clustering (NSC) agrees better with the NSC of the complete graph than the corresponding one obtained from PMFG. Finally, we show that the clusters in the PD network are more robust with respect to the removal of random sets of edges than those in the PMFG network.
\end{abstract}


\section{Introduction}

`Complex systems' is the term referring to the study of systems with a significant number of components in which we want to find out how the relationships between those components affect the behaviour of the system. The study of complex systems includes concepts from various disciplines such as mathematics, statistics, and computer science. One type of complex system is a complex network which consists of a large number of vertices and the relationships across them \cite{albert2002statistical}. There are many examples of such networks such as the Worldwide Web \cite{albert1999internet,barabasi2000scale}, papers citation network \cite{redner1998popular,small1973co}, social networks \cite{galaskiewicz1993social,wasserman1994social,watts2002identity,newman2002random}, and financial networks \cite{boss2004network,soramaki2007topology}.

One kind of financial network is a stock network. In such a network, the vertices denote the stocks and the weight of an edge between two stocks shows the similarity between them. Similarity could be, for example,  the influence the stocks have over the price of each other. One of the most commonly used measures to account for similarity in stock networks is Pearson correlation coefficient. That being said, including all the cross correlations across stocks would create a complete graph that reflects the complexity through a densely interwoven structure; because of which, several algorithms have been proposed to filter the complete graph into a simple subgraph to use as a representation of the original network. Some of these algorithms are minimum spanning tree (MST) \cite{mantegna1999hierarchical,bonanno2003topology,tabak2010topological,
guo2018development}, asset graph (AG) \cite{onnela2003asset}, planar maximally filtered graph (PMFG) \cite{tumminello2005tool,tumminello2007correlation,song2011evolution,wang2015correlation
,wang2017multiscale}, and correlation threshold method \cite{boginski2005statistical,huang2009network,chi2010network,namaki2011network}. Also, see Birch et al. \cite{birch2016analysis} for an advantages-and-limitations comparison of MST, AG, and PMFG on a dataset. Now the question is, what exactly makes a filtering algorithm better than the others? There is no unique answer, but to get a perspective, let us take a quick look at the reported positive aspects of the above-mentioned methods.

Mantegna \cite{mantegna1999hierarchical} attributed the advantage of MST to the fact that it provided a hierarchical clustering of stocks. Onnela et al. \cite{onnela2003asset} demonstrated the advantage of AG by observing that it had a higher survival ratio (ratio of common edges existing in two consecutive time steps) compared with MST. Yet they also mentioned that unlike MST, there was not an evident scale free behaviour indicating that the degree distribution follows a power law for AG. In sum, they found AG better in terms of being less fragile in the presence of market crisis, and that it incorporated more information from the original complete graph compared to MST, for it did not have the structural limitations of MST. Tumminello \cite{tumminello2005tool} attributed the usefulness of PMFG to the fact that the network produced always contains the one produced by MST, and that it contained cliques in it with stocks in those cliques mostly belonging to the same economic sectors. Boginski \cite{boginski2005statistical} mentioned that correlation threshold method was useful since for a large enough minimum threshold on the value of correlation coefficient, their network had a scale-free behaviour, and they could classify financial instruments through the analysis of cliques and independent sets of their network.
 
Others  have also discussed advantages of the above algorithms. Huang et al.~\cite{huang2009network} argued that coefficient threshold method displayed robustness against random vertex failures and a high average clustering coefficient. One of the points Wang et al. \cite{wang2017multiscale} made  is that PMFG is useful because it provided a good clustering of the stocks according to the economic sectoral benchmark clustering. In sum, the positive aspects of  filtering algorithms considered so far in the literature  are sparsity, scale-free behaviour, homogeneity of cliques, survival ratio, good clustering, and robustness. Of these, the clustering behaviour seems to attract the most attention.

We propose an algorithm called proportional degree (PD) to build a stock-correlation network based on the normalised mutual information (NMI) similarity matrix across the stocks. We show that the PD network with the same size as its PMFG counterpart has better homogeneity of cliques according to the stock economic sectors. We also show that the PD network has an overall better clustering compared to the PMFG network in terms of agreement with the normalised spectral clustering (NSC) of the similarity matrix.  

In Section \ref{method}, we define mutual information and explain why we used this measure to account for correlation across the stocks. Then we describe the  PD and PMFG algorithms in conjunction with the methods that we used to compare the corresponding networks of those algorithms. In Section \ref{results}, we provide the results of the comparison of the two networks built by the PD and PMFG algorithms. Finally, Section \ref{conclusion} includes our conclusion and some ideas for prospective researchers.

\section{Method} \label{method}

The measure we use to account for the correlation across stocks is NMI. The reason we prefer mutual information over correlation coefficient is that the former can detect the relationship between variables that cannot be detected by a linear correlation measure such as the latter \cite{da1989interdependence}. This feature of mutual information measure is more evident when the stock market exhibits violent fluctuations \cite{guo2018development}. We define this measure in the following.

Mutual information measures the level of independence between two random variables \cite{cover2012elements} where a value of zero shows statistical independence of the random variables. The mutual information between two stocks $X$ and $Y$ can be formulated as

\begin{equation}
\label{MIformula}
I\left( X,Y \right)= H(X)+H(Y)-H(X,Y)
\end{equation}
which is derived from Shannon's information entropy \cite{shannon2001mathematical}, a measure that quantifies the uncertainty of a random variable. Here, $I(X,Y)$ is the mutual information of $X$ and $Y$, $H(X)$ and $H(Y)$ denote entropy of $X$ and $Y$ respectively, and $H(X,Y)$ denotes the joint entropy of $X$ and $Y$. The entropy and joint entropy of discrete random variables $X$ and $Y$ are defined by 

\begin{equation}
\label{ent}
H(X)=-\sum\limits_{i}p(x_i)\log_2 p(x_i)
\end{equation}
\begin{equation}
\label{jointent}
H(X,Y)=-\sum\limits_i\sum\limits_jp(x_i,y_j)\log_2p(x_i,y_j)
\end{equation}
where $p(x_i)$ and $p(x_i,y_j)$ are the probability distribution and joint probability distribution of $X$ and $(X,Y)$ respectively. 

Unlike the correlation coefficient, mutual information is not bounded above by 1. Since large values of mutual information could be hard to interpret, it is useful to use the normalised mutual information, NMI, which  brings  the values down to the bounded interval  $ \left[0,1\right]$. It is defined by
\begin{equation}
\label{NMIformula}
\mathit{NMI}(X,Y)=\frac{2I(X,Y)}{H(X)+H(Y)}.
\end{equation}

One question we must face is how to construct the probability and joint probability distributions of the stocks in the {S\&P/ASX 200}
that we have chosen to study in order to find the mutual information between them? To this end, we use the same numerical method as proposed by Guo et al. \cite{guo2018development}. For $n$ stocks traded in $m$ business days, let $P_{it}$ be the closing price of stocks $i$ on day $t$. The log-return of stock $i$ on day $t$ for $t=2,3,\ldots,m$ and $i=1,2,\ldots,n$ is defined by
 
\begin{equation}
\label{log-return}
R_{it}=\ln{\frac{P_{it}}{P_{i(t-1)}}}.
\end{equation}

In order to find the probability distribution of the log-return of stock $i$, we sort $R_{it}$ values for $t=2,3,\ldots,m$ in ascending order and divide the sorted values into $q$ bins. Then we count the number of log-returns of stock $i$ for $i=1,2,\ldots,n$ in each bin $a$ for $a=1,2,\ldots,q$ denoted by $f_{ia}$ and get the approximate probability by $p_{ia}\approx\frac{f_{ia}}{m}$. Similarly, we find the joint probability distribution of the log-returns of stocks $i$ and $j$ for $i,j=1,2,\ldots,n$ by dividing their sorted log-returns into $q\times q$ bins. In such case, $f_{ijab}$ denotes the number of log-returns of $i$ and log-returns of $j$ in bin $(a,b)$, and the the approximate joint probability is given by $p_{ijab}\approx\frac{f_{ijab}}{m}$. As a result, we can approximate the entropy of stock $i$ and joint entropy of stocks $i$ and $j$ by 
\begin{equation}
\label{app_ent}
H(S_i)=-\sum\limits_{a=1}^q p_{ia} \log_2p_{ia}
\end{equation}
\begin{equation}
\label{app_joint_ent}
H(S_i,S_j)=-\sum\limits_{a=1}^q\sum\limits_{b=1}^q p_{ijab} \log_2 p_{ijab}.
\end{equation}
Therefore, the mutual information of stocks $i$ and $j$ can be given by substituting equations (\ref{app_ent}) and (\ref{app_joint_ent}) in equation (\ref{MIformula}), and the NMI is given by

\begin{equation} \label{NMI_matrix}
\mathit{NMI}(S_i,S_j)=\frac{2I(S_i,S_j)}{H(S_i)+H(S_j)} , \quad  i\neq j
\end{equation}
which produces a symmetric $n\times n$ matrix with diagonal elements of zero. We consider this matrix to be the similarity matrix of the stocks.

\subsection{Planar maximally filtered graph (PMFG)}

A graph can be represented by $G(V,E)$ in which $V=\{v_1,v_2,\ldots,v_n\}$ denotes the vertices and $E=\{e_{12},e_{13},\ldots,e_{ij},\ldots\}$ denotes the edges. A planar graph is one that can be embedded onto a surface with genus $g=0$, or the plane, without any two edges crossing or edges intersecting a vertex. The PMFG algorithm builds a network as follows.

\begin{algorithm}[H]
\renewcommand{\thealgorithm}{}
\caption{PMFG algorithm}\label{PMFG_algorithm}
\begin{algorithmic}[]
\State \textbf{Input:}
\State \quad  $V: \text{ set of stocks } $
\State \quad $s_{ij}: \text{similarity between stock $i$ and $j$ given in equation (\ref{NMI_matrix})} $
\State \textbf{Output:}
\State \quad $G(V,E): \text{planar network} $
\State \vspace{0.01cm}
\State $G(V,E) \gets \text{empty network of stocks \textit{V}} $
\State  $S\gets \mbox{ list of $(i,j,s_{ij})$ ($i,j \in V \, ,\, i \neq j)$, sorted in descending order}$  
\For {$ (i,j,s_{ij})$ in S }
\State $ E \gets E \cup \{e_{ij} \} $
\If {$G$ is planar $= False $}
\State $E \gets E - \{ e_{ij} \}$
\EndIf 
\EndFor
\end{algorithmic}
\end{algorithm}
The networks produced are maximal planar graphs and hence have $3n-6$ edges whenever $n$ is at least 3.

\subsection{Proportional degree (PD) algorithm}

We first determine the degree of each vertex in our output network in a manner such that it is proportional to its weight, where the weight of a vertex  (or stock weight) is the sum of its similarity value across all the other vertices. The weight of stock $i$ is defined by  
\begin{equation}
\label{stock_weight}
SW_i=\sum\limits_{j\neq i} s_{ij}
\end{equation}
where $SW_i$ and $s_{ij}$ respectively denote the weight of stock $i$ and the similarity between stocks $i$ and $j$.

Consequently, the calculated degree of a vertex ${d'_i}$ should be more or less given by

\begin{equation}
d'_i=\frac{SW_i}{\sum\limits_{j=1}^n SW_j} \times (2M)
\end{equation}
in which $M$ is the total number of edges, so $2M$ would be the sum of the degrees of all vertices. However, the degree of a vertex, being the number of adjacent vertices, is required to be integer. In order to round the calculated degrees $d_i'$ while preserving their total sum, we apply the  cascade rounding algorithm. For the rest of the paper, wherever we mention degree in association with the PD algorithm, it means the integer or rounded calculated degree. To use cascade rounding, we first relabel the vertices  as 1 to $n$, from largest stock weight to smallest.  Then we determine the degree of vertex $i$ recursively by subtracting the cumulative sum of the degrees of the $i-1$ vertices before it,  from the rounded cumulative sum of the calculated degrees of vertices 1 to $i$. Thus, $d_1=\nint{d'_1}$ and

\begin{equation}
\label{degrees}
d_i=\nint{\sum\limits_{j=1}^i d'_j}-\sum\limits_{j=1}^{i-1} d_j , \quad i\ge 2
\end{equation}
where $d_i$ is the degree of vertex $i$ and $\nint{x}$ denotes the nearest integer to $x$. Then the PD algorithm builds a network as follows.

\begin{algorithm}[H]
\renewcommand{\thealgorithm}{}
\caption{PD algorithm}\label{PD_algorithm}
\begin{algorithmic}[]
\State \textbf{Input:}
\State \quad  $V: \text{ set of stocks } $
\State \quad $s_{ij}: \text{similarity between stock $i$ and $j$ given in equation (\ref{NMI_matrix})} $
\State \textbf{Output:}
\State \quad $G(V,E): \text{proportional degree network} $
\State \vspace{0.01cm}
\State $G(V,E) \gets \text{empty network of stocks \textit{V}} $
\State   $S\gets \mbox{ list of $(i,j,s_{ij})$ ($i,j \in V \, ,\, i \neq j)$, sorted in descending order}$ 
\State $deg(i): \text{number of vertices adjacent to vertex \textit{i} in network \textit{G}} $ 
\For {$ (i,j,s_{ij})$ in S }
\If {$\left( deg(i) < d_i \right) $ and $\left( deg(j) < d_j \right)$  and 
$ \left( e_{ij} \notin E \right) $ }
\State $E \gets E \cup \{e_{ij}\} $
\EndIf 
\EndFor
\end{algorithmic}
\end{algorithm}

For the purpose of comparing with PMFG network, we set the total number of edges in this algorithm to $M=3n-6$ to equal the value in PMFG.

\subsection{Cliques} 
One of the advantages of PMFG over MST is the additional information linked with the inclusion of 3 and 4-cliques \cite{tumminello2005tool,tumminello2007correlation}. A clique is a subset of vertices in which every two vertices are connected via an edge. Such a subset is called a maximal clique if it is not contained in any larger clique. A clique of size $m$ is referred to as an $m$-clique. One way of analysing the cliques is to investigate how often the stocks in them belong to the same economic sector; in other words, what is the degree of cliques homogeneity with respect to the economic sectors \cite{tumminello2005tool}?  

\subsection{Clusters}
One of the most extensively investigated features of complex networks is community structure or clustering. Clusters in a graph are groups of vertices in which the density of edges inside those groups is considerably larger than the average edge density of the graph \cite{fortunato2010community}.  If each vertex of a graph only belongs to one cluster (no overlapping vertices), such a division of the graph determines a partition. Partitions of the stock-correlation PMFG networks have been widely studied \cite{buccheri2013evolution,song2011evolution, wang2015correlation, wang2017multiscale}. As with the analysis of cliques, one of the ways of analysing the clusters is investigating how well they match the economic sector classification of the stocks since we would hope that stocks belonging to the same economic sector are more likely to be in the same cluster \cite{chen2014analysis}.
We evaluate the clusters found by Louvain community detection \cite{blondel2008fast}, which is defined in the next subsection, in PD and PMFG networks through their similarity to the stocks' economic sectors partition. We also use Louvain community detection and normalised spectral clustering (NSC) \cite{shi2000normalized} later in the paper on the similarity matrix of the stocks (complete graph of NMI between stocks) and compare the resulting partitions with the partitions of the PD and PMFG networks achieved thorugh the same methods. To compare any two partitions, we use adjusted rand index (ARI) \cite{hubert1985comparing} which we discuss in more detail in subsection \ref{ARI_title}.

\subsubsection{Louvain community detection} \label{Louvain_subsection}

Louvain community detection is a greedy algorithm that tries to optimise the modularity function by choosing values $c_i$ for each vertex $i$ of the network. The modularity function is given as below. 

\begin{equation}
\label{modularity}
Q=\frac{1}{2S}\sum\limits_{ij} \left[s_{ij}-\frac{SW_i SW_j}{2S}\right] \delta(c_i,c_j)
\end{equation}
Here, $Q \in [-1,1]$, $S$ is the sum of all similarities (edge weights), $c_i$ and $c_j$ are the communities of stocks $i$ and $j$, $\delta$ is a simple delta function, and $SW_i$, $SW_j$, and $s_{ij}$ as already defined in equation (\ref{stock_weight}). In this algorithm, in the first step, each vertex is in its own community, that is, all the $c_i$'s are distinct. The effect on modularity caused by changing the community of a vertex $i$ to that of each of its neighbours in turn is checked. Then the community of vertex $i$ is reassigned to the community of the neighbour vertex that leads to the largest increase in modularity. In the case of no increase in modularity, $i$ keeps its own community label. This process is applied to all vertices and repeated until the community reassignment of none of the vertices leads to an increase in $Q$.  In the second step, all the the vertices belonging to the same community are considered as a single vertex, and the edges across vertices in the previous step are now denoted by self loops on the new vertex. Also, several edges from vertices of the same community in the previous step to a vertex in another community is denoted by a weighted edge between communities. These two steps are repeated iteratively until there is no change in the community assignment of the vertices in step one. 
That being said, depending on the order of vertices evaluated by this algorithm, we get different partitions. Accordingly, this algorithm does not yield the global maximum modularity. It is also worth mentioning  that finding the exact maximum modularity is an NP-hard problem, and one does not hope for an algorithm to solve it. 

\subsubsection{Normalised spectral clustering (NSC) } \label{NSC_subsection}

NSC is an algorithm that takes the similarity matrix and the number of clusters $k$ as its inputs and partitions the data set as below \cite{von2007tutorial,shi2000normalized}. 

\begin{algorithm}[H]
\renewcommand{\thealgorithm}{}
\caption{NSC algorithm}\label{NSC_algorithm}
\begin{algorithmic}[]
\State $W= {(w_{ij})}_{i,j=1,2,\ldots,n} : \text{similarity matrix} $ 
\State $k : \text{number of clusters}$
\State $ D : \text{digonal matrix with $d_i=\sum\limits_{j=1}^n w_{ij}$  , $i=1,2,\ldots,n  $ } $ 
\State $L=D-W : \text{Laplacian matrix} $
\State $\nu_1,\nu_2,\ldots,\nu_k \gets \text{eigenvectors of the $k$ smallest eigenvalues of the eigneproblem $L\nu=\lambda D \nu$}    $
\State $V \in \mathbb{R}^{n \times k} \gets \text{matrix with $\nu_1,\nu_2,\ldots,\nu_k$ as columns} $
\State $y_i \in \mathbb{R}^k \, , \, i=1,2,\ldots,n \gets \text{corresponding vector of $i$-th row of $V$ } $
\State $C_1,C_2,\ldots,C_k \gets \text{clusters of $y_i \in \mathbb{R}^k \, , \, i=1,2,\ldots,n$ by k-means algorithm} $  

\end{algorithmic}
\end{algorithm}

But what is a good choice of $k$? One tool to answer this question is eigengap heuristic \cite{von2007tutorial}. Defining sorted eigenvalues of Laplacian $L$ of the similarity matrix as $\lambda_1,     \lambda_2 , \ldots, \lambda_n$, eigengap heuristic states that the network should be divided into $k$ clusters so that $\lambda_{k+1}$ is significantly larger than $\lambda_1,\ldots,\lambda_k$. In other words, if the largest gap is between $\lambda_k$ and $\lambda_{k+1}$ for $k=1,2,\ldots,n-1$ in the sorted eigenvalues of the Laplacian of the similarity matrix, we divide the network into $k$ clusters.

\subsection{Adjusted Rand index (ARI)} \label{ARI_title}

The Rand index \cite{rand1971objective} is a measure in statistics that quantifies the similarity between two partitions of a data set. The ARI is another version of the Rand index that is corrected for chance. Given two partitions, namely $A$ and $B$ of the set $S$ containing $n$ elements, the ARI of $A=\{A_1,A_2,\ldots,A_r\}$ and $B=\{B_1,B_2,\ldots,B_s\}$ is as given by
\begin{equation}
\label{ARI}
\mathit{ARI}=\cfrac{\sum\limits_{ij} {\binom{n_{ij}}{2}} - \cfrac{\left[\sum\limits_{i} {\binom{a_i}{2}} \sum\limits_{j} {\binom{b_j}{2}} \right]}{{\binom{n}{2}}} }
{{\frac{1}{2} \left[ \sum\limits_{i} {\binom{a_i}{2}} + \sum\limits_{i} {\binom{a_i}{2}} \right]} - \cfrac{\left[\sum\limits_{i} {\binom{a_i}{2}} \sum\limits_{j} {\binom{b_j}{2}} \right]}{{\binom{n}{2}}}}
\end{equation}
where $n_{ij}=\vert A_i \cap B_j \vert$, $a_i=\sum\limits_{j=1}^s n_{ij} $, and $b_j=\sum\limits_{i=1}^r n_{ij} $. This measure satisfies $ARI \in [-1,1]$ so that 1 shows identical clusters, -1 shows complete mismatch, and 0 shows random assignment to clusters.

\section{Results} \label{results}
\subsection{Stock-correlation networks}
We selected 125 out of 200 stocks in the S\&P/ASX 200. The criterion used for selection was that these 125 stocks are the ones that were traded throughout the whole period of the years 2013-2016. The economic sectors and their number of corresponding stocks are as shown in the Table \ref{sectors_table}. 

\begin{table}[!htb]
  \centering
  \scriptsize
  \caption{Economic sectors of stocks}
  \renewcommand{\arraystretch}{1.5}
    \begin{tabular}{|l|l|}
    \hline
    \multicolumn{1}{|c|}{\textbf{Economic Sector}} & \multicolumn{1}{c|}{\textbf{Number of Stocks}} \\
    \hline \hline
    Consumer Dscretionary & 21 \\
    \hline
    Consumer Staples & 6 \\
    \hline
    Energy & 8 \\
    \hline
    Financials & 19 \\
    \hline
    Health Care & 10 \\
    \hline
    Industrials & 16 \\
    \hline
    Information Technology & 2 \\
    \hline
    Materials & 26 \\
    \hline
    Real Estate & 12 \\
    \hline
    Telecommunication Services & 2 \\
    \hline
    Utilities & 3 \\
    \hline
    \end{tabular}%
  \label{sectors_table}%
\end{table}%

In order to get the NMI between all the stocks to generate the similarity matrix, we chose the bin size of $q=20$ (referring to equations (\ref{app_ent}) and (\ref{app_joint_ent})) for the 1013 trading days in our data since as Guo et al. \cite{guo2018development}  mention,   for a large enough $q$, there is not much difference in the values of mutual information, and they considered a bin size of $q=10$ for the 734 trading days in their data.  

\begin{figure}[!htb]
\centering
	\begin{subfigure}[]{\textwidth}
		\centering
		\includegraphics[width=0.65\columnwidth]{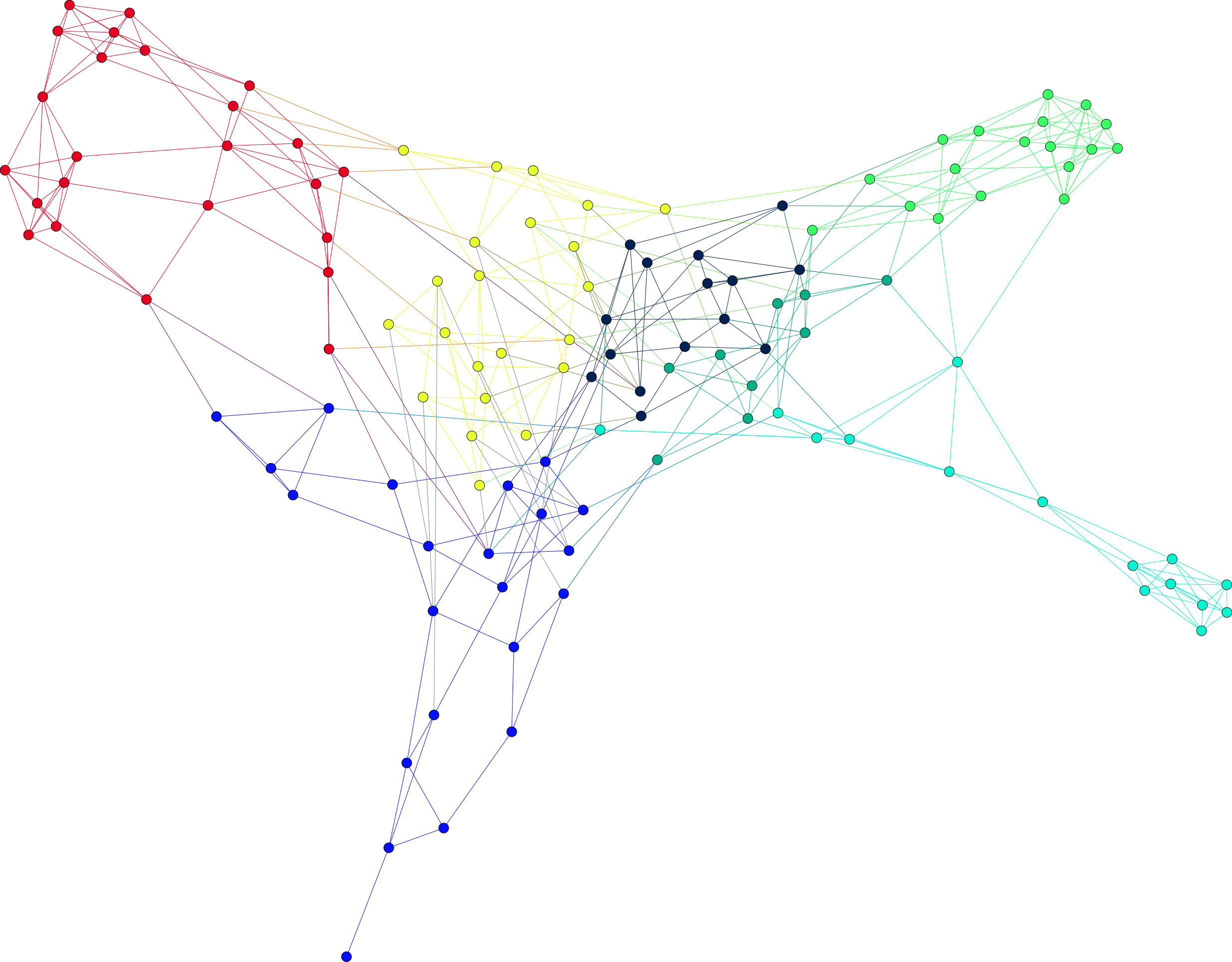}
		\caption{Different colors referring to different clusters of Table \ref{PD clusters} 				}
	\end{subfigure}
	\vskip\baselineskip
	\begin{subfigure}[]{\textwidth}
		\centering
		\includegraphics[width=0.65\columnwidth]{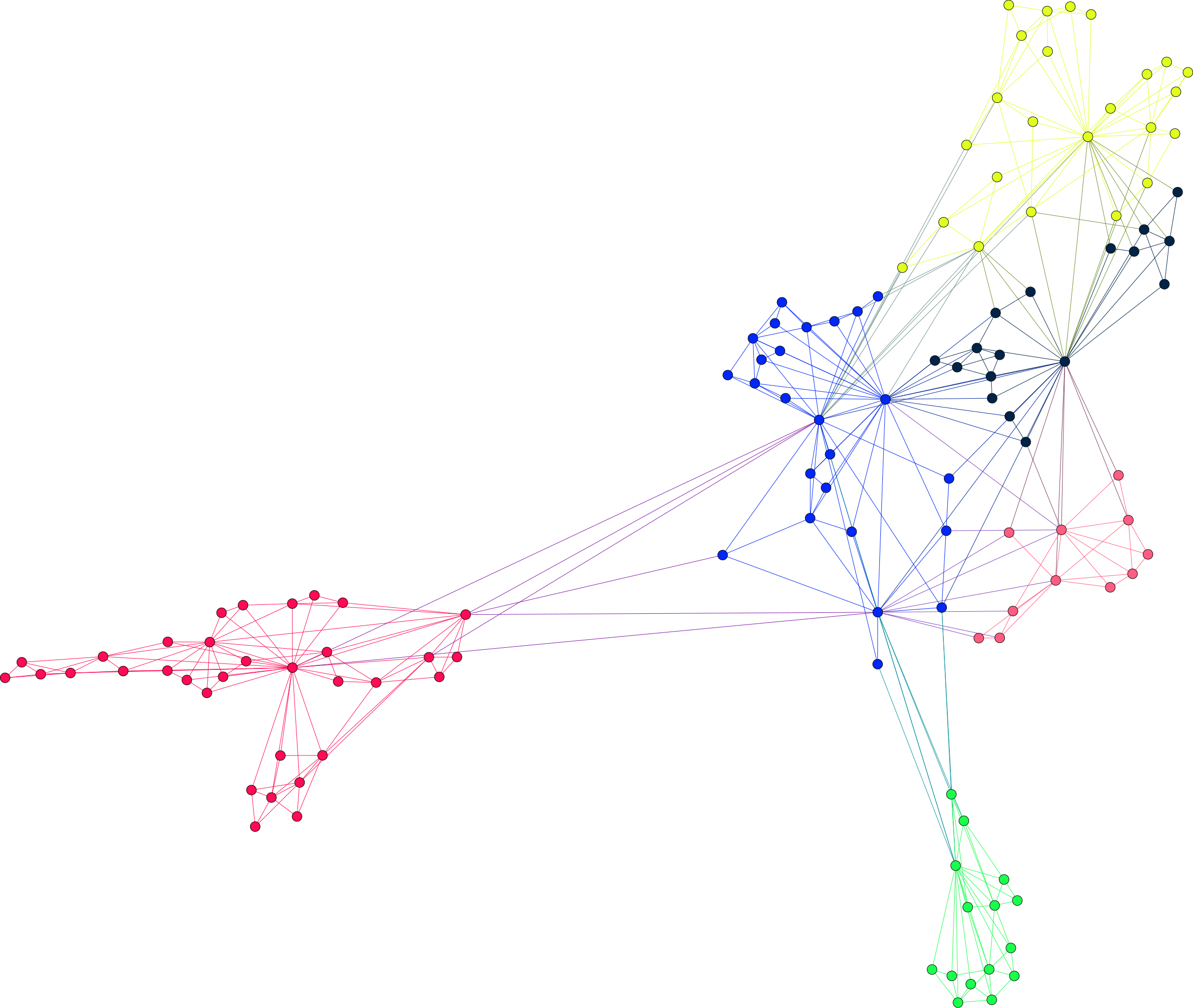}
		\caption{Different colors referring to different clusters of Table \ref{PMFG clusters} 				}
	\end{subfigure}
\caption{Clusters found in the PD (a) and PMFG (b) networks using Louvain community 				detection}
\label{networks_visual}
\end{figure}

We generated the PD and PMFG networks for the above-mentioned data, the analysis of which is provided below. In the PD network, we have vertices with degrees ranging from 1 to 9 
whereas in the PMFG network the degrees range from 3 to 29. Visualisations of both networks are in Figure \ref{networks_visual}.


\subsection{Cliques}

\begin{figure}[!htb]
\centering
\includegraphics[width=0.7\columnwidth]{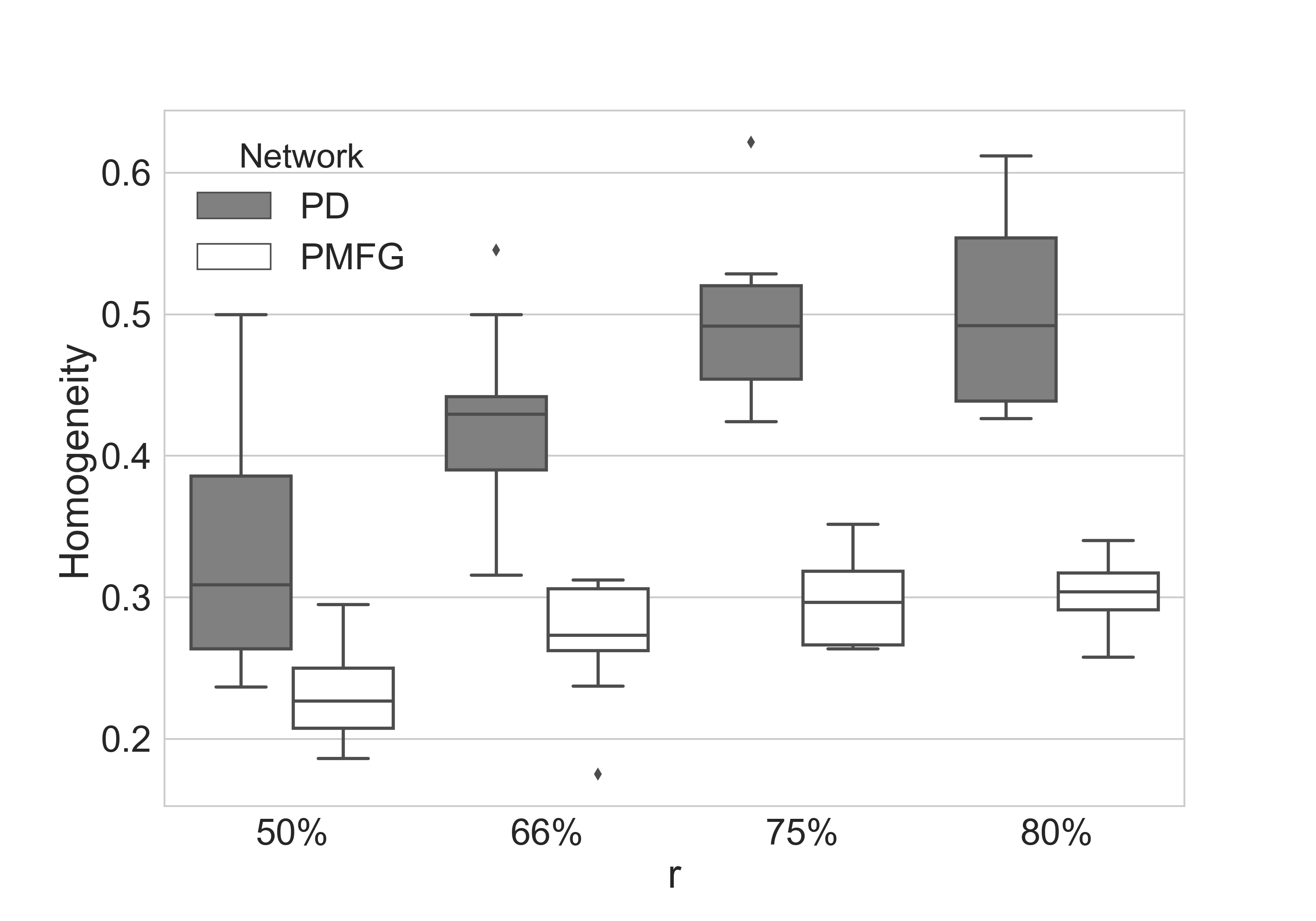}
\caption{Maximal cliques homogeneity comparison of the two networks on different random subsets of proportions $r$ of all stocks}
\label{homogeneity}
\end{figure}

The analysis of the networks generated in the previous subsection indicates that there are 87 maximal cliques of size 3 and larger, including 52 maximal cliques of size 3, 23 of size 4, 9 of size 5, and 3 of size 6. Similarly, there are 122 maximal cliques of size 4 in PMFG network. To quantify homogeneity, $\nicefrac{47}{87}=0.54$ of the maximal cliques in the PD network consist of stocks all belonging to the same economic sector whereas this ratio is $\nicefrac{43}{122}=0.35$ for the PMFG network. We also compared the homogeneity of the maximal cliques with minimum size of 3 in the two networks on different random subsets of the stocks. To this end, we considered different proportions $r=\nicefrac{4}{5}, \nicefrac{3}{4}, \nicefrac{2}{3}, \nicefrac{1}{2} $ of all the 125 stocks, and for each $r$, we took 10 samples of size $\nint{r \times n}$ from the stocks. We plotted the results of all samples for each $r$ and each network as shown in Figure \ref{homogeneity}, and we can see that for every $r$, the PD network has an overall larger homogeneity of maximal cliques compared with PMFG.

\begin{figure}[!htb]
\centering
	\begin{subfigure}[]{0.475\textwidth}
		\centering
		\includegraphics[width=\textwidth]{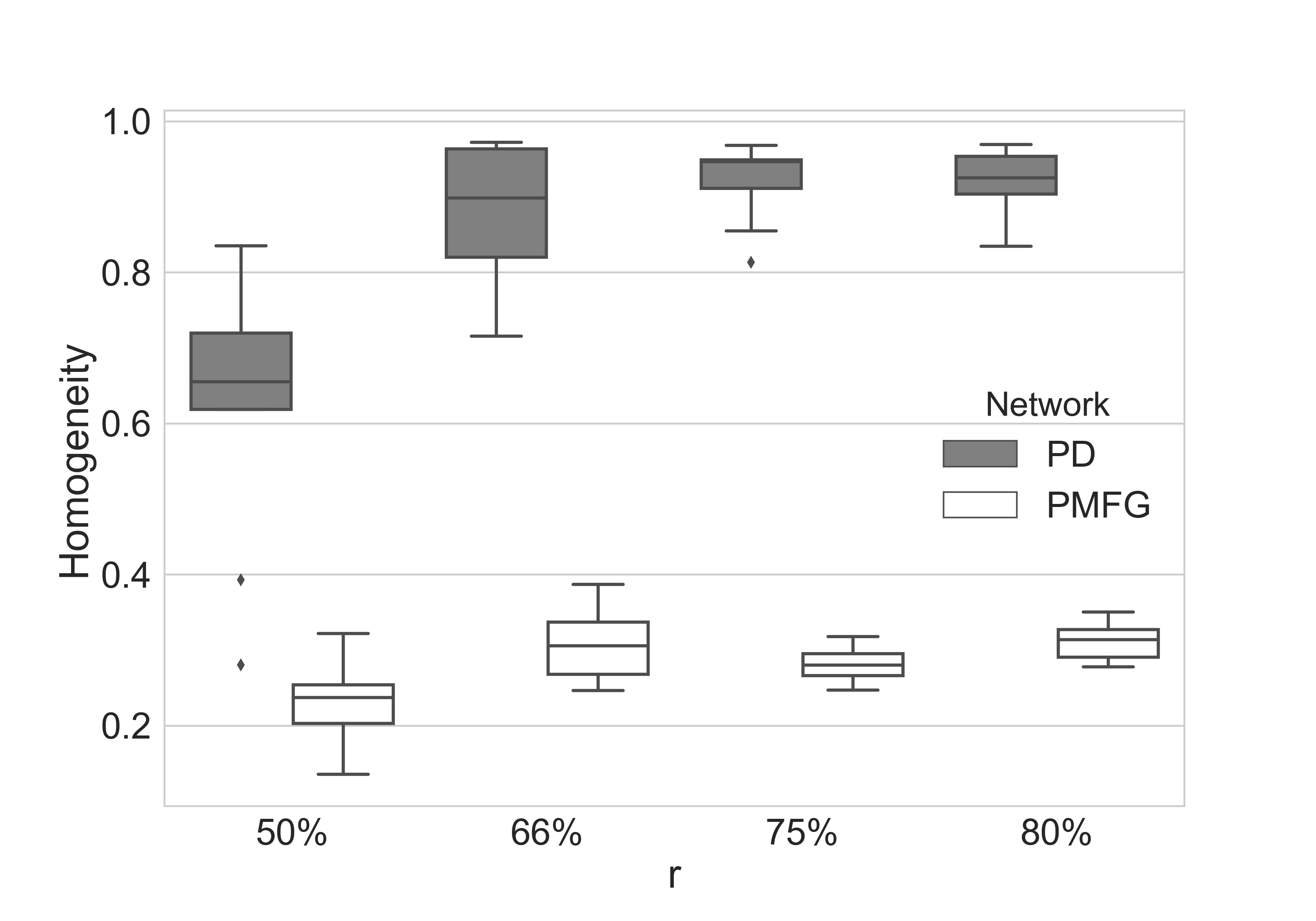}
		\caption{4-cliques}
	\end{subfigure}
	\hfill
	\begin{subfigure}[]{0.475\textwidth}
		\centering
		\includegraphics[width=\textwidth]{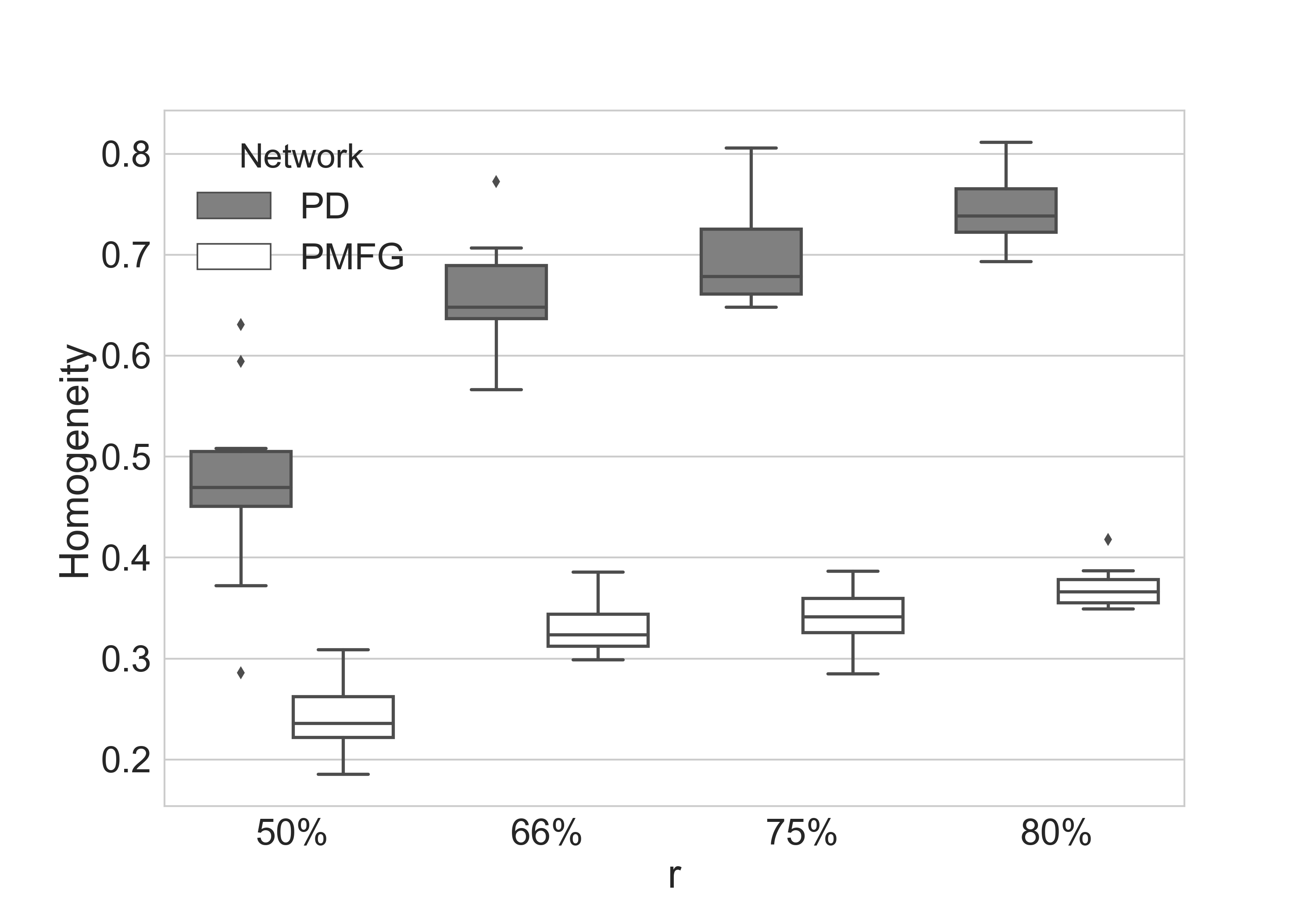}
		\caption{3-cliques}
	\end{subfigure}
\caption{4-cliques and 3-cliques homogeneity comparison of the two networks on different random subsets of proportions $r$ of all stocks}
\label{homogeneity_2}
\end{figure}

Contrasting with this, in a PMFG network, as any other planar graph, we cannot have a maximal clique with size more than 4 since such a clique cannot be embedded into a surface with genus $g=0$ without any two edges crossing. In fact, we can have at most $n-3$ maximal 4-cliques and $3n-6$ 3-cliques in any planar graph \cite{wood2007maximum}. Accordingly, in another analysis, we compared the homogeneity of the 3-cliques and 4-cliques in the PD network, which are not necessarily maximal, with their counterparts in the PMFG network. There are 236 3-cliques and 101 4-cliques in the PD network, and we observe a homogeneity of $\nicefrac{178}{236}=0.75$ in the 3-cliques and $\nicefrac{91}{101}=0.90$ in the 4-cliques of it. These ratios are $\nicefrac{152}{367}=0.41$ and $\nicefrac{43}{122}=0.35$ in the PMFG network. So we can also see that all the maximal cliques in the PMFG network are 4-cliques here. As with the previous analysis, we compared the homogeneity of 3-cliques and 4-cliques of the two networks on different random subsets of the stocks, and the result is plotted on Figure \ref{homogeneity_2}. We can see that the comparison of 3-cliques and 4-cliques homogeneity between the two networks is even more striking than for maximal cliques.

\subsection{Clusters}

Using the Louvain community detection approach, following Wang and Xie\cite{wang2015correlation} and Wang et al. \cite{wang2017multiscale}, we identified clusters as shown in Tables \ref{PD clusters} and \ref{PMFG clusters}. 
We found 7 clusters in PD and 6 in PMFG. In both tables, ``Sectors'' refers to the number of stocks belonging to each economic sector in the corresponding cluster, e.g. 17F denotes 17 stocks belonging to Financials sector, ``Size'' refers to the cluster size, ``Dominant'' refers to the economic sector repeated the most in the cluster, and ``Percentage'' refers to the proportion of the stocks belonging to the dominant sector in the cluster. 

\begin{table}[!htb]
  \centering
  \scriptsize
  \caption{Clusters captured in the PD network by Louvain community detection}
  \label{PD clusters}
   \renewcommand{\arraystretch}{1.5}
    \begin{tabular}{|c  H | p{10em} |l|l|l|}
    \hline
    \small{Cluster} & \small{Cluster} & \small{Sectors} & \small{Size}  & \small{Dominant } & \small{Percentage} \\
    \hline \hline
    1     & CBA, WBC, ANZ, NAB, AMP, SUN, MQG, IAG, ASX, LLC, BEN, BOQ, PTM, CGF, IFL, HGG, PPT, MFG & 17F, 1RE & 18    & F     & 94\% \\
    \hline
    2     & BHP, WPL, RIO, ORG, FMG, STO, OSH, ORI, WOR, ILU, ALQ, AWC, SVW, BSL, WHC, DOW, MND, SGM, MIN, BPT, OZL, SFR, IGO, WSA & 13M, 7E, 4I & 24    & M     & 54\% \\
    \hline
    3     & TLS, TCL, GMG, SGP, GPT, SYD, MGR, APA, IOF, CQR, BWP, CHC, ABP, SCP, MQA & 10RE, 3I, 1U, 1TS & 15    & RE    & 67\% \\
    \hline
    4     & WOW, WES, CSL, BXB, RHC, SHL, COH, PRY, IVC & 5HC, 2CS, 1CD, 1I & 9 & HC  & 56\% \\
    \hline
    5     & NWS, NCM, RMD, DXS, QAN, DUE, TPM, SKI, ANN, NVT, RRL, MSB, TME, FXJ, MMS, EVN, SIP, GWA, SAI, SRX, GUD, NST & 5CD, 5HC, 4I, 4M, 2U, 1RE, 1TS & 22    & CD    & 23\% \\
    \hline
    6     & QBE, AMC, CCL, AZJ, CTX, CPU, AIO, IPL, FBU, JHX, BLD, TWE, MTS, GNC, ABC, SWM, MYR, SXL, CSR, NUF, PBG, AAD & 8M, 5CD, 4CS, 2I, 1E, 1F, 1IT & 22 & M & 36\% \\
    \hline
    7     & CWN, TTS, SEK, FLT, HVN, SUL, TAH, ALL, DLX, QUB, JBH, PMV, FXL, IRE, BRG & 10CD, 2I, 1F, 1IT, 1M & 15    & CD    & 67\% \\
    \hline
    \end{tabular}%
\end{table}%

\begin{table}[!htb]
  \centering
  \scriptsize
  \caption{Clusters captured in the PMFG network by Louvain community detection}
  \label{PMFG clusters}
  \renewcommand{\arraystretch}{1.5}
    \begin{tabular}{|c H | p{10em} |l|l|l|} 
    \hline
    {\small{Cluster}} &  \small{Cluster} & \small{Sectors} & \small{Size} & \small{Dominant} & 			\small{Percentage} \\ [0.5ex]
    \hline \hline
   {1}     & CBA, ANZ, NAB, TLS, WOW, WES, QBE, SUN, IAG, CCL, AZJ, TTS, BEN, MTS, BOQ, GNC, TPM, TAH, CGF, MSB, DLX, QUB, FXL, GWA, AAD & 10F, 5CS, 3CD, 3I, 2TS, 1HC, 1M & 25    & F     & 40\% \\
    \hline
    2     & BHP, WPL, RIO, NCM, ORG, FMG, STO, OSH, ORI, WOR, CTX, AIO, IPL, ILU, ALQ, AWC, SVW, BSL, WHC, DOW, MND, SGM, RRL, MIN, BPT, OZL, FXJ, NUF, EVN, SFR, IGO, WSA, NST & 19M, 8E, 5I, 1CD & 33    & M     & 58\% \\
    \hline
    3     & WBC, LLC, FBU, JHX, QAN, BLD, HVN, SUL, ABC, SWM, MYR, JBH, PMV, SXL, CSR, PBG, GUD & 9CD, 5M, 1F, 1I, 1RE & 17    & CD    & 53\% \\
    \hline
    4     & CSL, RMD, RHC, SHL, DXS, COH, TWE, PRY, ANN, SIP, SRX & 9HC, 1CS, 1RE & 11    & HC    & 82\% \\
    \hline
    5     & NWS, AMP, BXB, MQG, AMC, CWN, ASX, CPU, SEK, FLT, PTM, ALL, NVT, IFL, HGG, PPT, TME, IVC, MMS, MFG, IRE, BRG, MQA, SAI & 8CD, 8F, 5I, 2IT, 1M & 24    & CD    & 33\% \\
    \hline
    6     & TCL, GMG, SGP, GPT, SYD, MGR, APA, DUE, SKI, IOF, CQR, BWP, CHC, ABP, SCP & 10RE, 3U, 2I & 15    & RE    & 67\% \\
    \hline
    \end{tabular}
\end{table}%

Nonetheless, as pointed out in Section \ref{Louvain_subsection}, Louvain community detection yields different partitions depending on the order of vertex evaluation. To mitigate the effect of different partitions corresponding to different orders of vertices on ARI, we applied the Louvain method on 100 random orders of vertices in both networks and took the average of those 100 ARIs for each method in terms of resemblance to economic sectors' partition of stocks. This produced average ARIs of 0.31 and 0.26 for PD and PMFG networks respectively.

However, the economic sector classification is not the be-all and end-all partition of stocks. For example, every stock labelled as Real Estate in the ASX/S\&P 200 data of 01/10/2018 had been put in the Financials category in the ASX/S\&P 200 data of 21/03/2016,  which means that the economic sector classification is subject to change, reducing the likelihood that it represents the unique correct partition. 
Indeed, it could also
be argued that there are some significant sub-categories in other economic sectors, which would create more clusters than the number of economic sectors. To create another partition benchmark  other than the economic sector classification, we used Louvain community detection on the complete graph of NMI between stocks (similarity matrix of the stocks). This computation produced only four  clusters of stocks. Comparing clusters achieved by Louvain community detection in PD and PMFG networks as shown in Tables \ref{PD clusters} and \ref{PMFG clusters} with the new partition benchmark, we got ARIs of 0.40 and 0.36 respectively. Yet 
not much can be concluded from this comparison since the number of clusters in the benchmark partition is so different from the numbers of clusters in the two networks.

\begin{figure}[!htb]
\centering
\includegraphics[width=0.7\columnwidth]{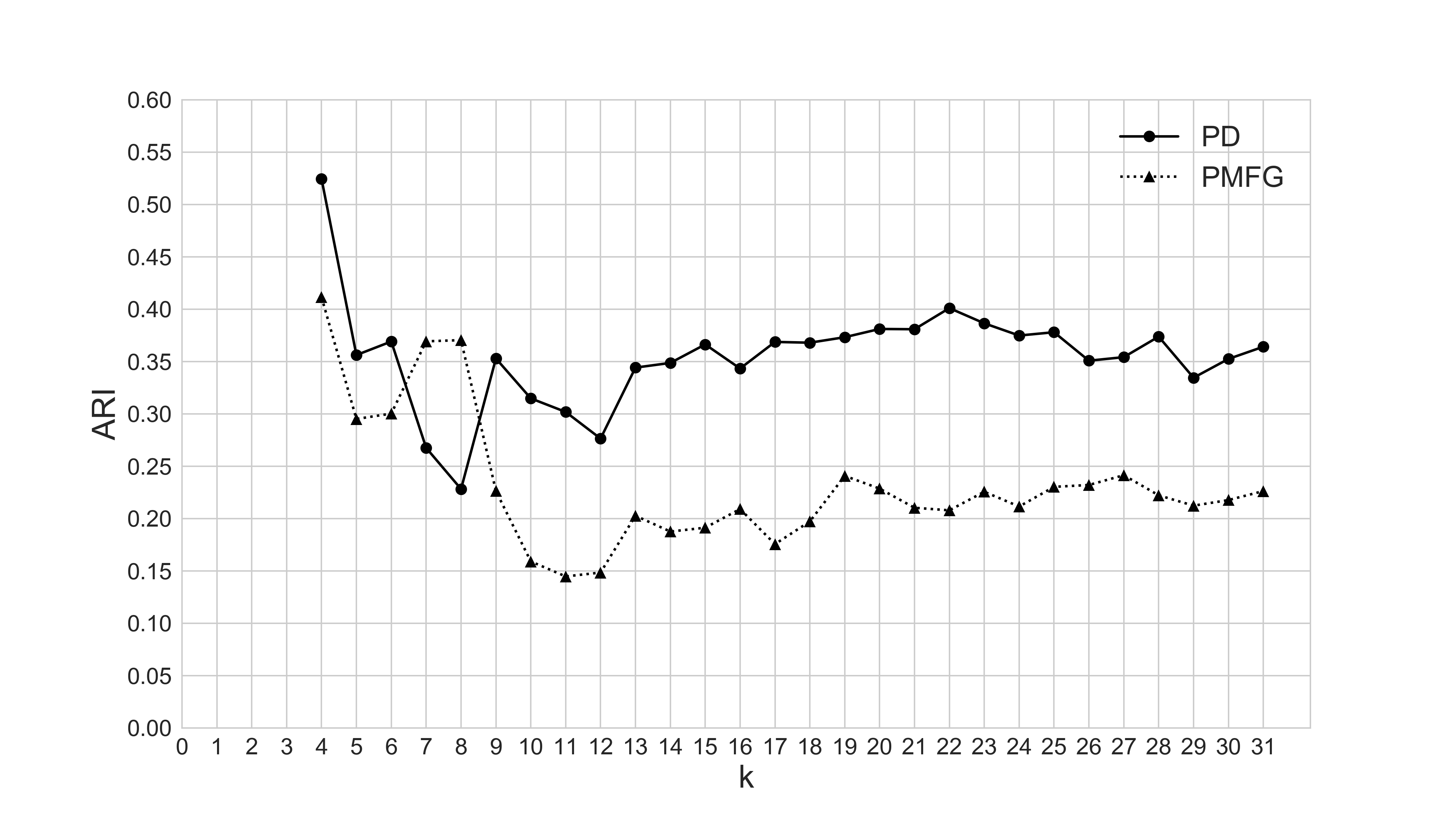}
\caption{ARI performance comparison of $C_{\textrm {PD}}$ versus $C_{\textrm {PMFG}}$}
\label{r=1}
\end{figure}

To  draw a more significant comparison between the clustering behaviour of the two networks, we used NSC on the similarity matrix of the stocks and called the resulting partition $C_{\textrm K}$. Then we applied NSC to the PD and PMFG networks where the corresponding partitions are denoted by $C_{\textrm{PD}}$ and $C_{\textrm{PMFG}}$ respectively. 
For the similarity matrix to input into NSC, we used the binary adjacency matrix of the networks.  
In Figure \ref{r=1}, the Y-axis denotes the ARI of $C_{\textrm K}$ and $C_{\textrm {PD}}$ versus that of $C_{\textrm K}$ and $C_{\textrm{PMFG}}$, and the X-axis denotes $k$. Here, we regard a  network to have a good ARI performance if its ARI against $C_{\textrm {K}}$ is large. It can be seen that for small values of $k$, there is not much difference in the ARI performance of the networks, for $k=7,8$, PMFG has a better ARI performance, and for $k>8$, PD consistently has a better ARI performance than PMFG.  As shown in Figure \ref{r=1}, we restrict the number of clusters to being at least 4 because this is the least number of clusters in the application of Louvain to any of the networks or graphs under discussion, and is much smaller than the number of economic sectors. 

Implementing the heuristic described in Section \ref{NSC_subsection} and ignoring the gaps between the first, second, and third sorted eigenvalues since we ignore 1 and 2 as the number of clusters, we find the largest gaps between the sorted eigenvalues for the similarity matrix are $g(\lambda_4,\lambda_5)=0.74$, $g(\lambda_{10},\lambda_{11})=0.35$, and $g(\lambda_{11},\lambda_{12})=0.16$. We then note that the PD network has a better ARI performance than the PMFG network for $k=4,10,11$. 
From another perspective, one of the points we made is that there could be some subsectors lurking in the classification of stocks by the economic sector. As we have 11 economic sectors, from this point of view, the number of clusters can be $k>11$, and for these values of $k$, PD consistently displays a better ARI performance than PMFG. It should be said that although spectral clustering does not perform well on sparse networks all the time \cite{krzakala2013spectral,le2015sparse,amini2013pseudo}, NSC gives a sensible result in our networks as the partitions match fairly well with the economic sector classification $C_e$ of the stocks as shown in Table \ref{NSC vs economic sectors}. Also, the result of this table is another indicator that small values of $k$ are not valid, for the ARI of $C_{\textrm {PD}}$ and $C_e$ is smaller than that of $C_{\textrm K}$ and $C_e$. Besides this, the ARI of $C_{\textrm {PD}}$/$C_{\textrm {PMFG}}$ and $C_e$ is small for small values of $k$ compared to larger values of $k$.

\begin{table}[!htb]
	\scriptsize
  \centering
  \caption{ARI of $C_{\textrm {PD}}$/$C_{\textrm {PMFG}}$/$C_{\textrm K}$ and $C_e$}
  \renewcommand{\arraystretch}{1.5}
    \begin{tabular}{|l|l|l|l|}
    \hline
    \multicolumn{1}{|p{4.215em}|}{\textbf{k}} & \multicolumn{1}{p{4.215em}|}{\textbf{PD}} & \multicolumn{1}{p{4.215em}|}{\textbf{PMFG}} & \multicolumn{1}{p{7.5em}|}{\textbf{Complete graph}} \\
    \hline \hline
    5     & 0.195 & 0.0585 & 0.121 \\
    \hline
    6     & 0.197 & 0.0921 & 0.124 \\
    \hline
    7     & 0.195 & 0.069 & 0.229 \\
    \hline
    8     & 0.236 & 0.1665 & 0.27 \\
    \hline
    9     & 0.339 & 0.1557 & 0.352 \\
    \hline
    10    & 0.242 & 0.1015 & 0.376 \\
    \hline
    11    & 0.274 & 0.0833 & 0.29 \\
    \hline
    12    & 0.279 & 0.0799 & 0.33 \\
    \hline
    \end{tabular}%
  \label{NSC vs economic sectors}%
\end{table}%

\begin{figure}[!htbp]
\centering
	\begin{subfigure}[]{0.475\textwidth}
		\centering
		\includegraphics[width=\textwidth]{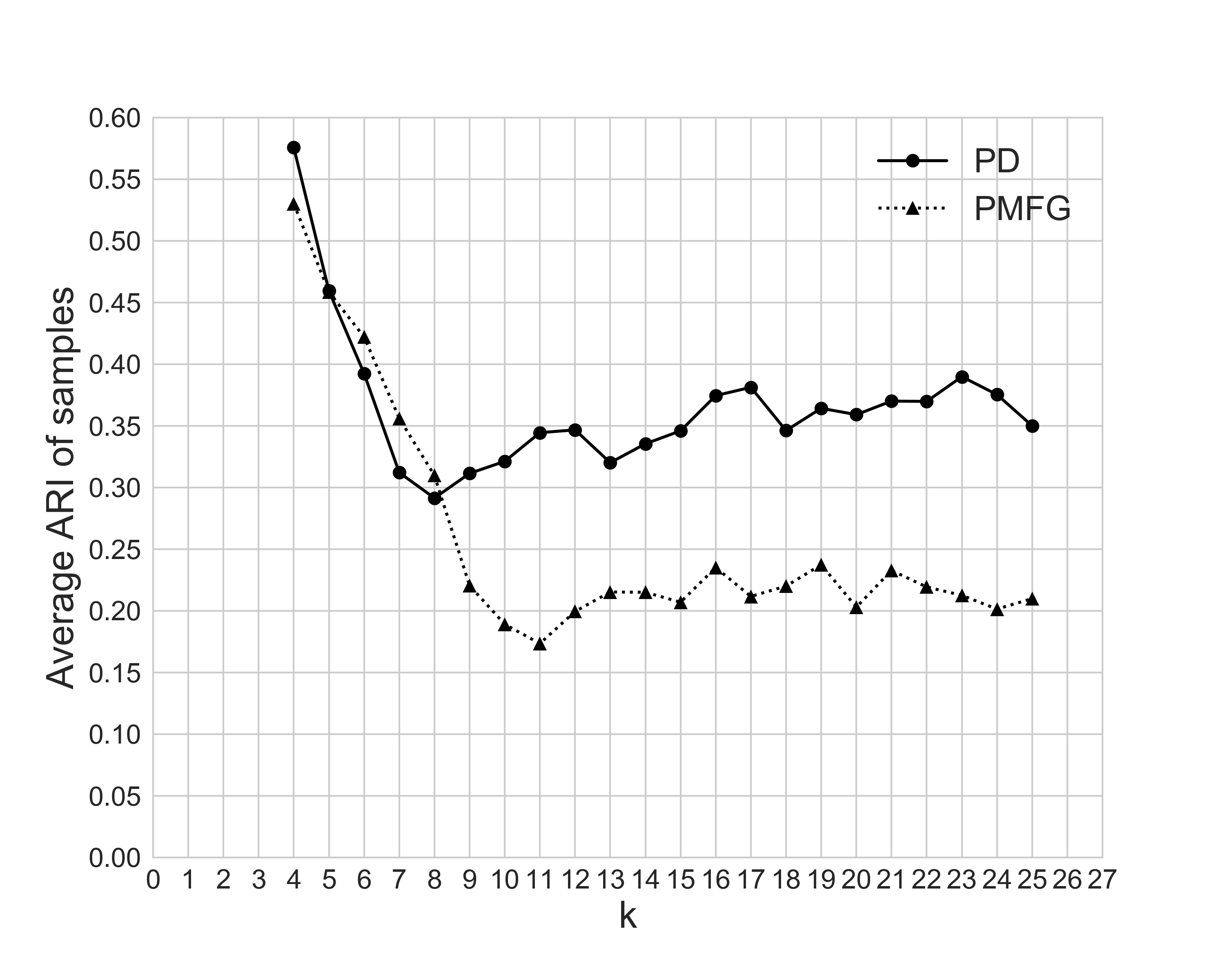}
		\caption{$r=\nicefrac{4}{5}$}
	\end{subfigure}
	\hfill
	\begin{subfigure}[]{0.475\textwidth}
		\centering
		\includegraphics[width=\textwidth]{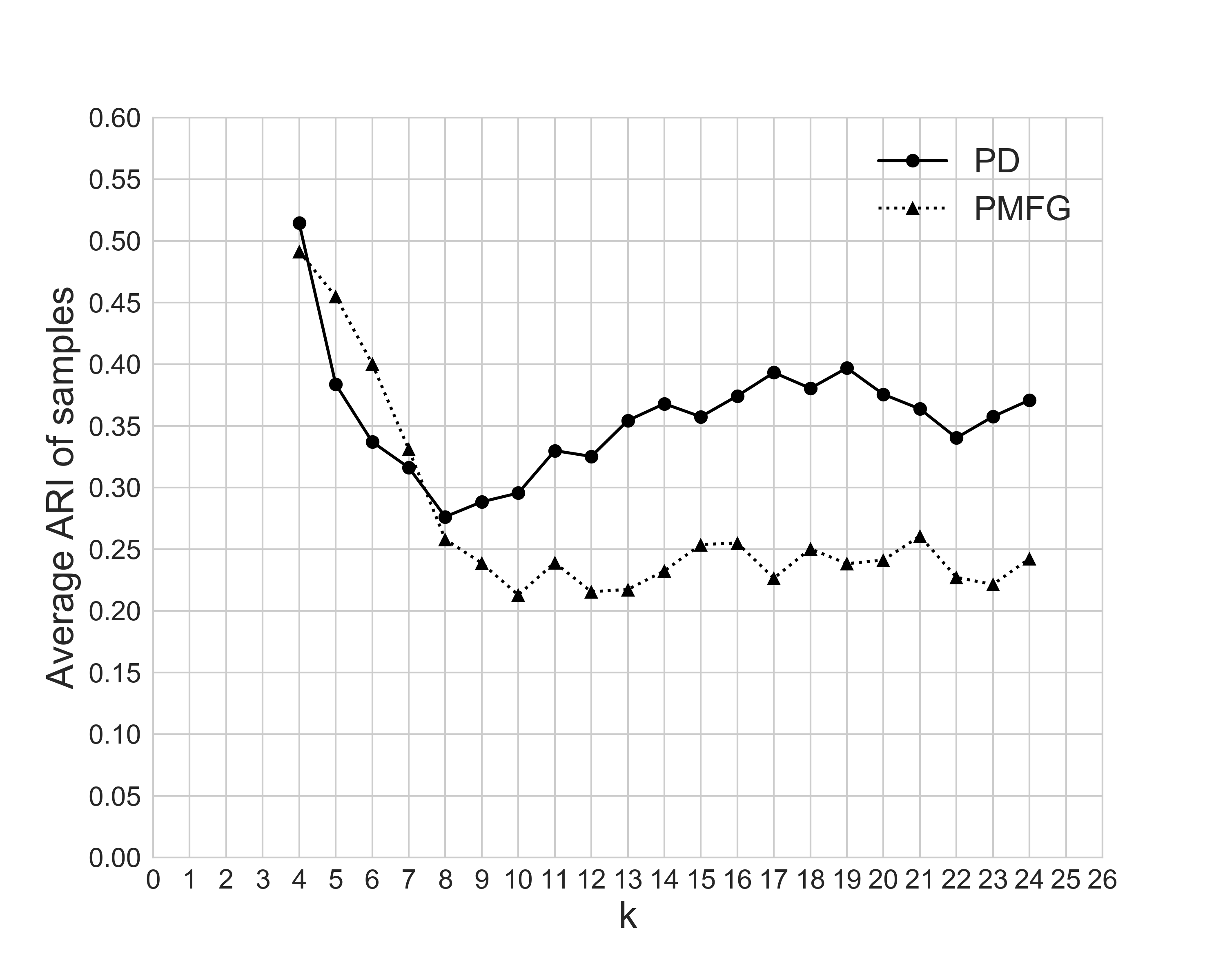}
		\caption{$r=\nicefrac{3}{4}$}
	\end{subfigure}
	\vskip\baselineskip
	\begin{subfigure}[]{0.475\textwidth}
		\centering
		\includegraphics[width=\textwidth]{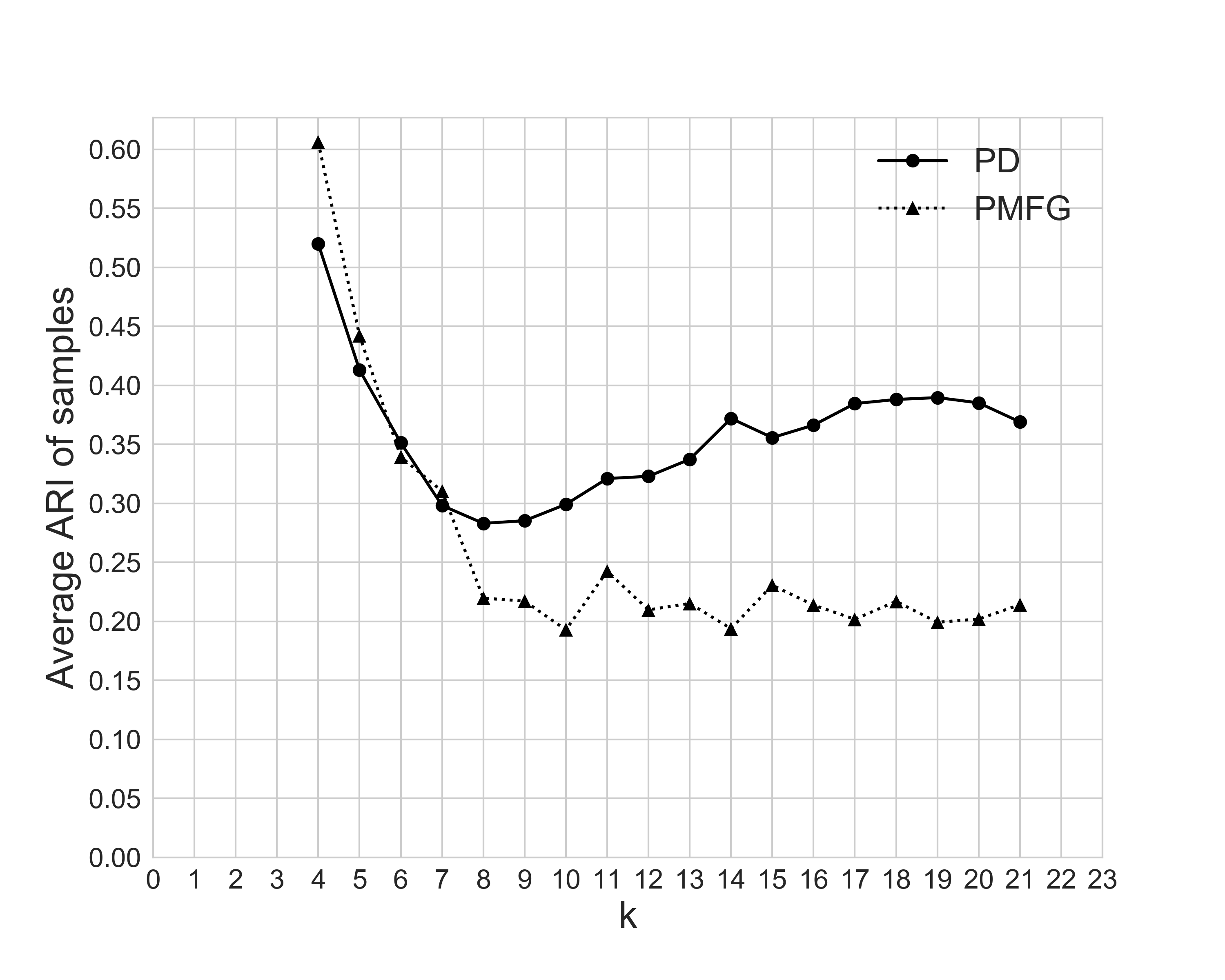}
		\caption{$r=\nicefrac{2}{3}$}
	\end{subfigure}
	\hfill
	\begin{subfigure}[]{0.475\textwidth}
	\centering
		\includegraphics[width=\textwidth]{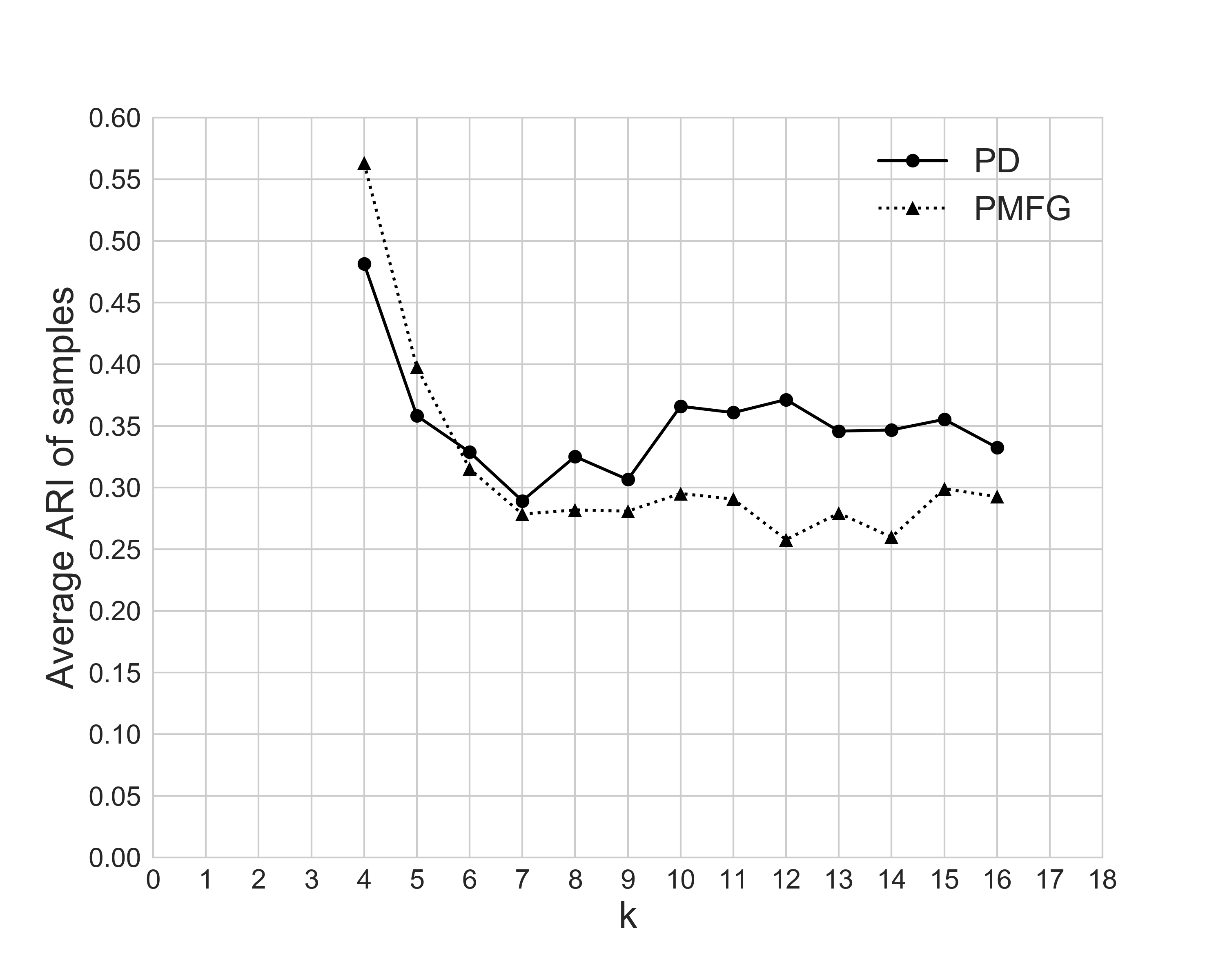}
		\caption{$r=\nicefrac{1}{2}$}
	\end{subfigure}
\caption{Average ARI performance of the PD and PMFG networks for different proportions $r$ of stocks}
\label{different ratios}
\end{figure}

As with our analysis of the homogeneity of cliques, to test the validity of our result, we also compared the ARI performance of the two networks on different random subsets of the stocks. To this end, again, we considered different possible proportions $r=\nicefrac{4}{5}, \nicefrac{3}{4}, \nicefrac{2}{3}, \nicefrac{1}{2} $, and for each $r$, we took 10 samples of size $\nint{r n}$ from the stocks. Then we implemented the PD and PMFG algorithms on the samples to generate the two networks and applied NSC on both networks for each sample. Denoting the partitions of PD, PMFG, and the complete similarity matrix of sample $i$ by $C_{{\rm{PD}}_i} $, $C_{{\rm{PMFG}}_i} $, and $C_{\rm{K}_i}$ respectively, we considered the average ARI of $C_{\rm{K}_i}$ and $C_{{\rm{PD}}_i}$  versus that of $C_{\rm{K}_i}$ and $C_{{\rm{PMFG}}_i} $ for $i=1,\ldots,10$
and plotted the results as shown in Figure \ref{different ratios}. We can see the same pattern for every $r$; that for a large enough $k$, PD has consistently a better average ARI performance than PMFG whereas for small values of $k$, there is virtually no difference in the average ARI performance of the networks. In addition, we can see on Figure \ref{different ratios} that as the size of the network shrinks (for smaller values of $r$), the difference between the average ARI performance of the networks become smaller. In other words, forcing into a planar network has less effect on clustering of the stocks in smaller networks. One reason could be that there is less use of larger cliques in smaller networks; thus, the restriction of PMFG to maximal clique size of 4 becomes less important. 

\subsection{Robustness}

\begin{figure}[!htbp]
\centering
	\begin{subfigure}[]{\textwidth}
		\centering
		\includegraphics[width=0.7\columnwidth]{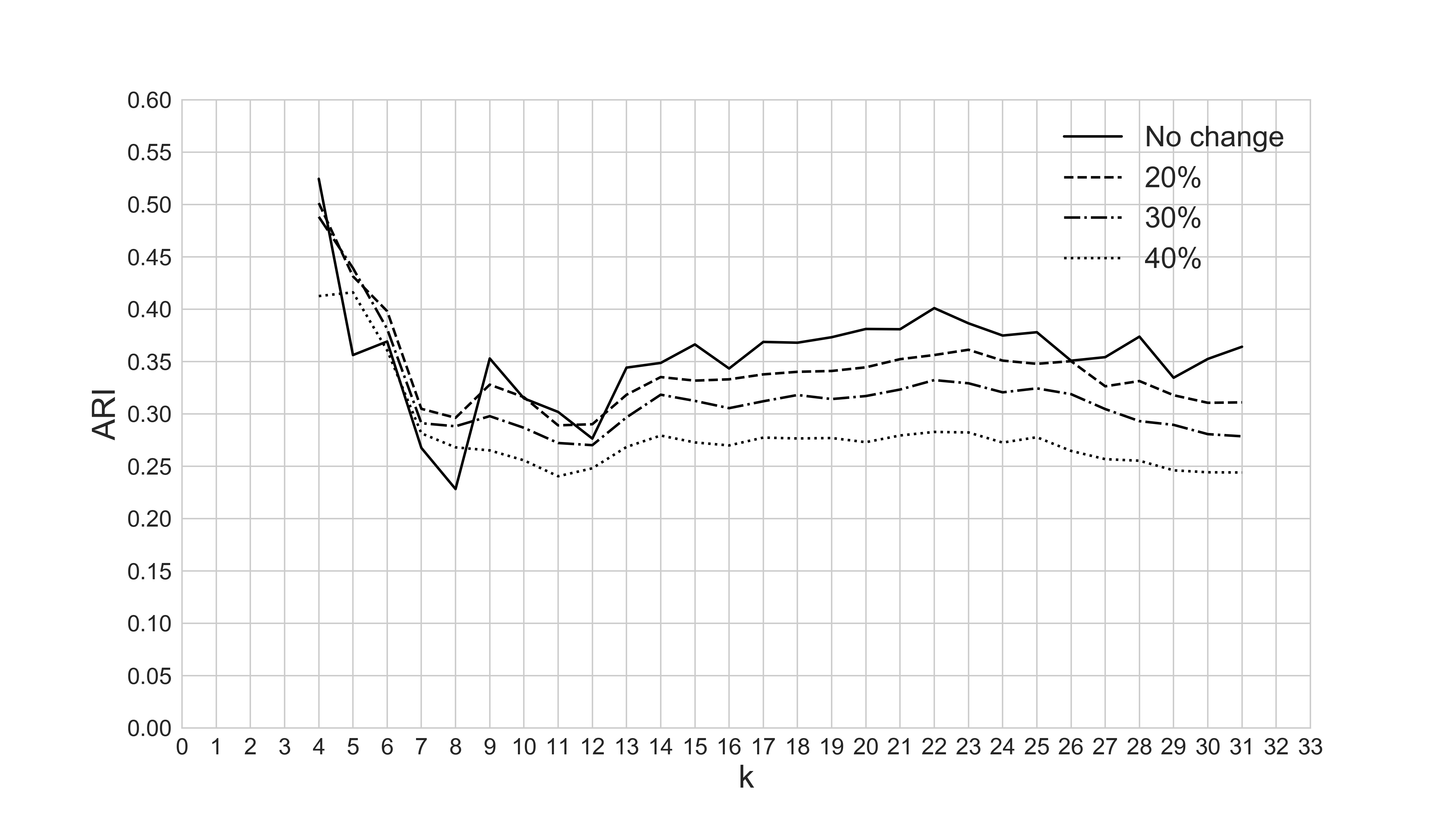}
		\caption{PD}
	\end{subfigure}
	\vskip\baselineskip
	\begin{subfigure}[]{\textwidth}
		\centering
		\includegraphics[width=0.7\columnwidth]{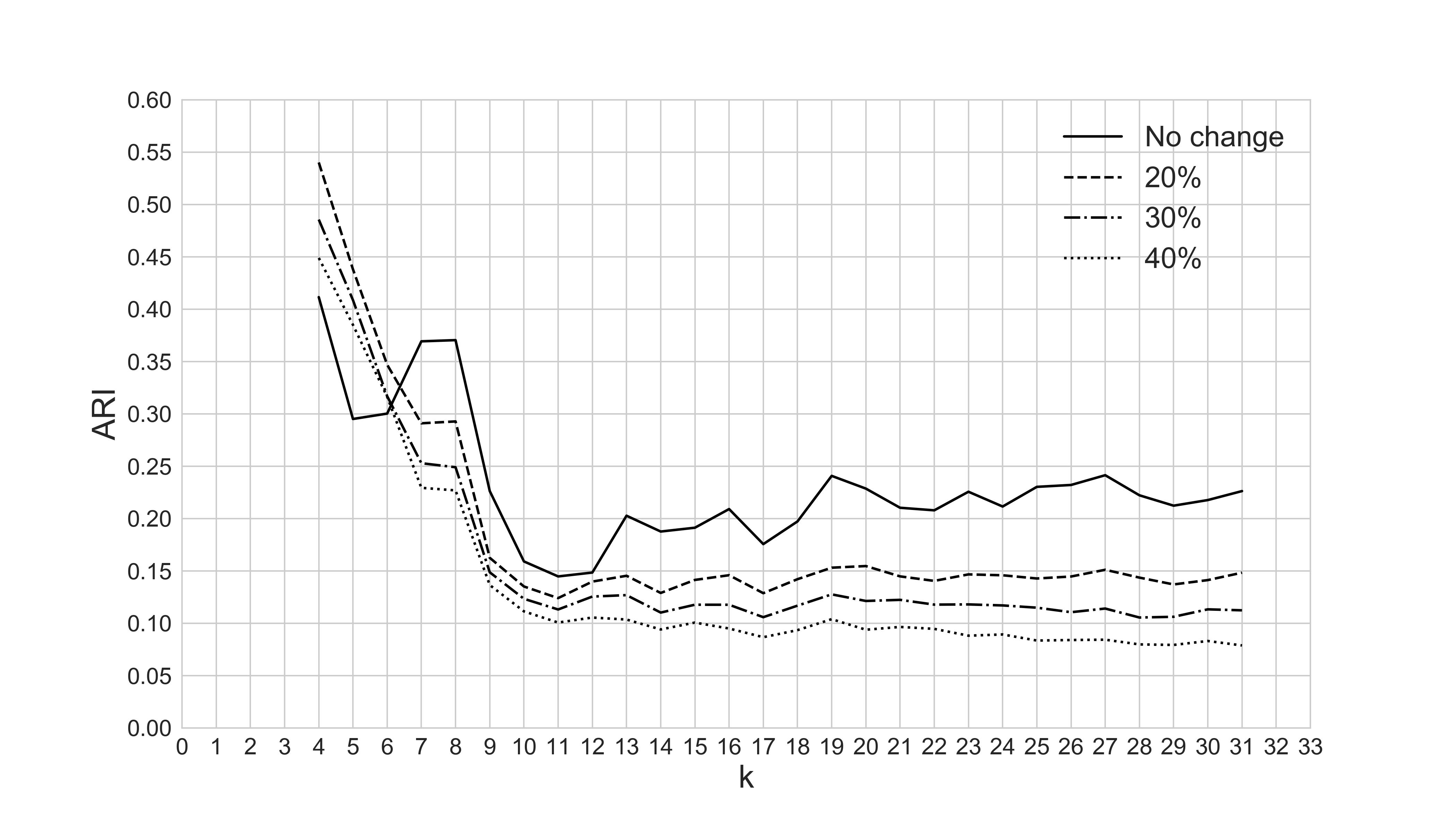}
		\caption{PMFG}
	\end{subfigure}
\caption{Fluctuations in ARI for NSC of the networks for different proportions of edge removal}
\label{clusters_robustness}
\end{figure}

One method to investigate the robustness or stability of network is removing a subset of its vertices or edges at a certain rate \cite{huang2009network}. On both networks, we removed 100 different samples of 20\%, 30\%, and 40\% of the edges randomly and applied NSC on them. Then we plotted the the average ARI of $C_{\rm{K}}$ and $C_{\rm{PD}}$ and that of $C_{\rm{K}}$ and $C_{\rm{PMFG}}$  as shown in Figure \ref{clusters_robustness}. As expected, there is an overall decrease in ARI performance of both networks as the percentages of edge removal increases. That being said, there is an increase in the average ARI for small $k$'s ( $k\leq 8$ and $k\leq 6$ for the PD and PMFG networks respectively), which could be another indicator that small values of $k$ are not valid. Hence, PD has a better clustering behaviour that PMFG since for large values of $k$, it displays a better ARI performance.

\begin{figure}[!htbp]
\centering
\includegraphics[width=0.7\columnwidth]{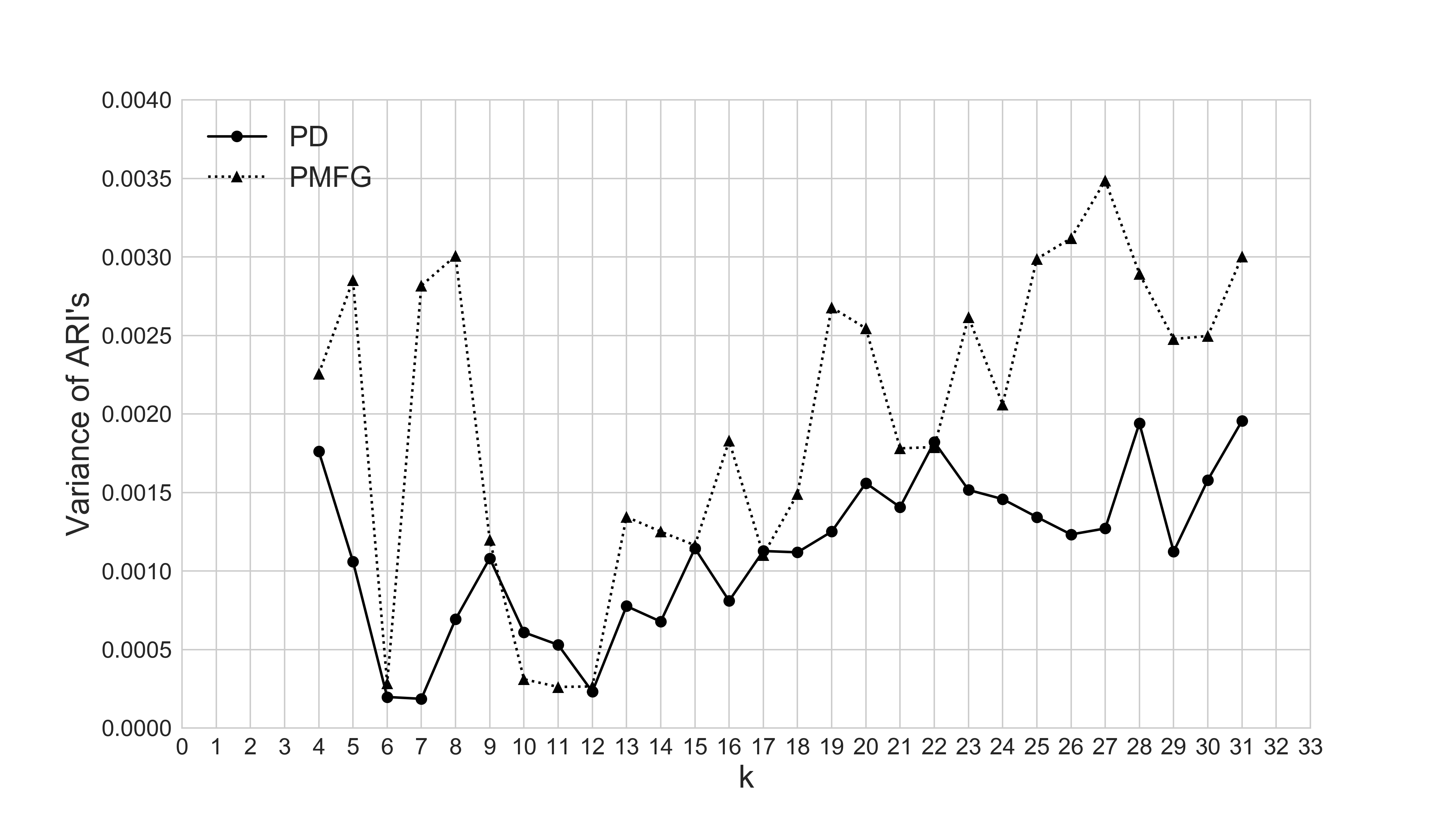}
\caption{Robustness of the networks clusters in presence of edge removal for each $k$ }
\label{ARIs_variance}
\end{figure}

In order to see which network has more change of clusters by edge removal, we took the variance of ARIs of both networks for each $k$ in 4 states, being firstly the networks with no change, and then the networks with 20\%,30\%, and 40\% edge removal respectively. The results are plotted on Figure \ref{ARIs_variance}, and we can see that for every $k$, there is either not a significant difference in the variance of the ARIs or PD has a significantly smaller variance than PMFG; thus, more robust with respect to change in clusters. 


\FloatBarrier

\section{Conclusion} \label{conclusion}
We used the NMI measure to build a cross-correlation similarity matrix across stocks and applied the PD and PMFG algorithms to generate the corresponding stock-correlation networks. We showed that maximal cliques, 3-cliques, and 4-cliques had a higher homogeneity in the PD network than the PMFG network as to financial sectoral classification of the stocks. Moreover, we showed that for a realistic number of clusters in the NSC algorithm, the PD network has a better ARI performance than the PMFG network in terms of matching the clusters achieved through applying the NSC algorithm on the similarity matrix of the stocks. 

It should be noted that the aforementioned results were achieved using NMI, and they are not necessarily expected using other correlation measures. Also, we used $3n-6$ edges to build the PD network for the whole purpose of comparing its performance with its PMFG counterpart. It is not clear this size of the PD network is the optimal one considering the criteria of a superior stocks-correlation network. A future topic for prospective researchers can be varying the sparsity of the PD algorithm and comparing the resulting networks. Also, other measures of correlation and dependence across stocks such as Spearman's rank correlation coefficient could be used to build the stock-correlation network and compare that with other stock-correlation networks according to the criteria used in the literature.

\bibliographystyle{unsrt}
\bibliography{method}
\clearpage

\end{document}